\documentclass[10pt,final,conference,letterpaper,twocolumn]{IEEEtran}

\usepackage[utf8]{inputenc}
\usepackage[bookmarks=false]{hyperref} 
\usepackage{flushend} 
\usepackage{xcolor,color,colortbl,graphicx,tabularx}
\usepackage{amssymb, pifont}
\usepackage{etoolbox}
\usepackage[para]{threeparttable}
\usepackage{pgfplots}
\usepackage{multirow}
\usepackage{xspace}
\usepackage{cite}
\usepackage[acronym]{glossaries}
\usepackage{booktabs}
\usepackage{todonotes}
\presetkeys%
{todonotes}%
{inline}{}
\usepackage{comment}

\usepackage{nicefrac}
\usepackage{xfrac}
\usepackage{subfloat}
\usepackage{subcaption}

\usepackage{tikz}
\usetikzlibrary{shapes,arrows,shadows}
\usepackage{amsmath,bm,times}
\usepackage{environ}

\usepackage[group-minimum-digits=3,group-separator={,}]{siunitx}

\makeatletter
\newsavebox{\measure@tikzpicture}
\NewEnviron{scaletikzpicturetowidth}[1]{%
	\def\tikz@width{#1}%
	\begin{lrbox}{\measure@tikzpicture}%
		\BODY
	\end{lrbox}%
	\pgfmathparse{#1/\wd\measure@tikzpicture}%
	\BODY
}

\newcommand{\gl}{geolocation\xspace}
\newcommand{\drop}{DRoP\xspace}
\newcommand{\hloc}{HLOC\xspace}

\newcommand{\rtt}{round trip time\xspace}

\newcommand{\eg}{\textit{e.g.,}\xspace}
\newcommand{\ie}{\textit{i.e.,}\xspace}

\newcommand{\zmap}{ZMap\xspace}
\newcommand{\ripeatlas}{RIPE Atlas\xspace}

\colorlet{tablelight}{black!15}
\colorlet{tabledark}{black!40}

\title{\hloc: Hints-Based Geolocation Leveraging\\ Multiple Measurement Frameworks}

\author{\IEEEauthorblockN{Quirin Scheitle, Oliver Gasser, Patrick Sattler, Georg Carle}
	\IEEEauthorblockA{Chair of Network Architectures and Services\\
        Technical University of Munich (TUM)\\
		Email: \{scheitle,gasser,sattler,carle\}@net.in.tum.de}
}

\begin{document}
\maketitle
%
\begin{abstract}
Geographically locating an IP address is of interest for many purposes. 
There are two major ways to obtain the location of an IP address: querying commercial databases or conducting latency measurements.
For structural Internet nodes, such as routers, commercial databases are limited by low accuracy, while current measurement-based approaches overwhelm users with setup overhead and scalability issues.\\
In this work we present our system \hloc, aiming to combine the ease of database use with the accuracy of latency measurements.
We evaluate \hloc on a comprehensive router data set of 1.4M IPv4 and 183k IPv6 routers.
 \hloc first extracts location hints from rDNS names, and then conducts multi-tier latency measurements.
Configuration complexity is minimized by using publicly available large-scale measurement frameworks such as \ripeatlas. 
Using this measurement, we can confirm or disprove the location hints found in domain names.
We publicly release \hloc's ready-to-use source code, enabling researchers to easily increase \gl accuracy with minimum overhead. 
\end{abstract}
\section{Introduction}\label{sec:intro}%
Geographical location of IP addresses has many purposes and users. 
Public service and businesses frequently focus on locating persons through their end-user devices. 
These human-centric use cases typically rely on easy-to-use commercial databases, which provide ever-increasing accuracy for edge nodes through feedback from users affected by wrong entries. 
For academia, the location of structural Internet nodes such as routers is elemental in studying and mapping the Internet. 
Unfortunately, commercial databases often provide low quality in this use case. 
State-of-the-art measurement-based algorithms are often not ready to use, but require implementation and setup of measurement infrastructure before use. 
For these reasons, structural Internet studies today frequently use commercial databases, and try to alleviate errors of these databases through laborious outlier analysis\,\cite{feamster2016detours, schmitt2016mvno, scheitle2016analyzing}.
To cope with these shortcomings, we present \hloc, a framework combining the simplicity and scale of commercial \gl databases with the accuracy of measurement-based approaches. 
The basic idea of \hloc is to leverage location hints frequently observed in router DNS names \cite{chabarek2013name, huffaker2014drop}.
As these hints may be incorrect or ambiguous \cite{spring2002measuring, zhang2006misnaming}, \hloc interfaces with ready-to-use measurement frameworks to confirm or invalidate those hints through latency measurements. 
With the geographic coverage of frameworks such as \ripeatlas~\cite{ripeatlas}, it is usually possible to find a probe very close to a location candidate and potentially confirm a location hint within a 100km radius.
While a 100km radius would be considered city-level \gl, \hloc may also return a bigger radius of confidence, which may be considered country-level \gl.
This radius of confidence can be configured by the user to match specific use cases.
For the large scale of this work, which aims to locate all IPv4 and IPv6 router addresses found in CAIDA's traces, we precede this measurement step by integrating high-volume \zmap~\cite{zakirzmap,Gasser2016ipv6} scans to (i) filter for responsive hosts and (ii) quickly locate the hemisphere of an IP address.
This pre-measurement step can be skipped if only few IP addresses need to be located.
We release HLOC source code and data used in this paper to the public.
As \ripeatlas automates data sharing between different measurements, the repeated usage of \hloc by multiple research teams will further improve its efficiency.

\noindent Our main contributions in this work are:
\begin{itemize}
    \item Creating \hloc, a ready-to-use \gl framework leveraging DNS-based hint collection and distributed active measurements
    \item In-depth evaluation of \hloc \gl measurements of 1.4M IPv4 and 183k IPv6 routers
    \item Comparison of \hloc \gl results against commercial databases and previous measurement-based approaches
\end{itemize}

We structure this paper as follows: We discuss related work in Section \ref{sec:relwork}  and present \hloc's architecture in Section \ref{sec:arch}. 
We discuss measurement results in Section \ref{sec:measres} and evaluate our results against other approaches in Section \ref{sec:eval}. 
We discuss limitations, trade-offs, future improvements and ethical considerations in Section \ref{sec:disc}. 
Section \ref{sec:concl} concludes this work. 
\section{Related Work}\label{sec:relwork}
Related work on \gl can be structured into three groups: measurement-based approaches, DNS-based approaches, and investigations on accuracy limitations of \gl databases.

\textbf{Measurement-based: }%
There is a large body of measurement-based approaches for IP address \gl.
These approaches rely on latency, link level topology, time-to-live and other data to triangulate hosts or co-locate hosts to known landmarks\, \cite{Hu2012TGM,wang2011towards,wong2007octant,katz2006towards,gueye2006constraint,youn2009statistical,yoshida2009inferring, guo2009mining, padmanabhan2001investigation,calder2013mapping}.

\textbf{DNS-based: }%
In contrast to measurement-based approaches, DNS-based approaches leverage DNS data to derive an IP address's location.
RFC\,1876 defines a DNS extension capable of storing longitude/latitude data for IP addresses; however, this extensions is rarely used, and would likely suffer from outdated information when IP addresses are being moved without updated DNS entries.

Several previous works use DNS data to locate IP addresses. Spring et al.\,\cite{spring2002measuring} leverage DNS data to locate Internet routers for a topology study. B\"ottger et al. \cite{bottger2016open} study Netflix's infrastructure in depth by extracting information from DNS names and confirming those by measurements. 
Zhang et al.\,\cite{zhang2006misnaming} show that misnamed router DNS names (e.g. due to re-assignment of IP addresses without changing the DNS name) can significantly distort topological studies.
They suggest various rules to detect such misnaming, for example by disallowing location loops along a traceroute's path.

In 2014, Huffaker et al.\,\cite{huffaker2014drop} (``\drop'') investigate router DNS names and compare them against various ground truth domains and databases. %
They also use CAIDA's Internet measurement infrastructure to derive location rules for domains through latency and time-to-live analysis.
While showing good results on few ground truth domains, \drop does not provide a ready-to-use solution as its source code is not public, it leverages private measurement frameworks, and its generalization potential is unclear.
In this work, we reappraise \drop using our \gl framework \hloc by (i) including more hint sources; (ii) using a more flexible search algorithm; (iii) leveraging public measurement frameworks; (iv) adding IPv6 capabilities and (v) publishing ready-to-use code and data.%

\textbf{Database Accuracy: }%
Gueye et al.\,\cite{gueye2007investigating} in 2007 criticize the concept of \gl databases to aggregate IP addresses into larger blocks. They show that some of these blocks can span 1,000km and more.
Poese et al.\,\cite{poese2011ip} in 2011 investigate the accuracy of five different \gl databases against ground truth data for one large Internet Service Provider. They find the databases to be partially incorrect on a city level, and heavily biased on a country level.
They argue that this inconsistency is likely increased by block aggregation as criticized by Gueye et al.\,\cite{gueye2007investigating}.
We strive to remediate accuracy problems of \gl databases with \hloc by providing a ready-to-use \gl framework with higher accuracy.

\section{Architecture of HLOC}\label{sec:arch}
We design HLOC as a modular, flexible and extensible framework.
This allows us to scale efficiently and accommodate changing requirements. 

\subsection{\hloc's Building Blocks}

\begin{figure}
	\centering
	\includegraphics[width=\columnwidth]{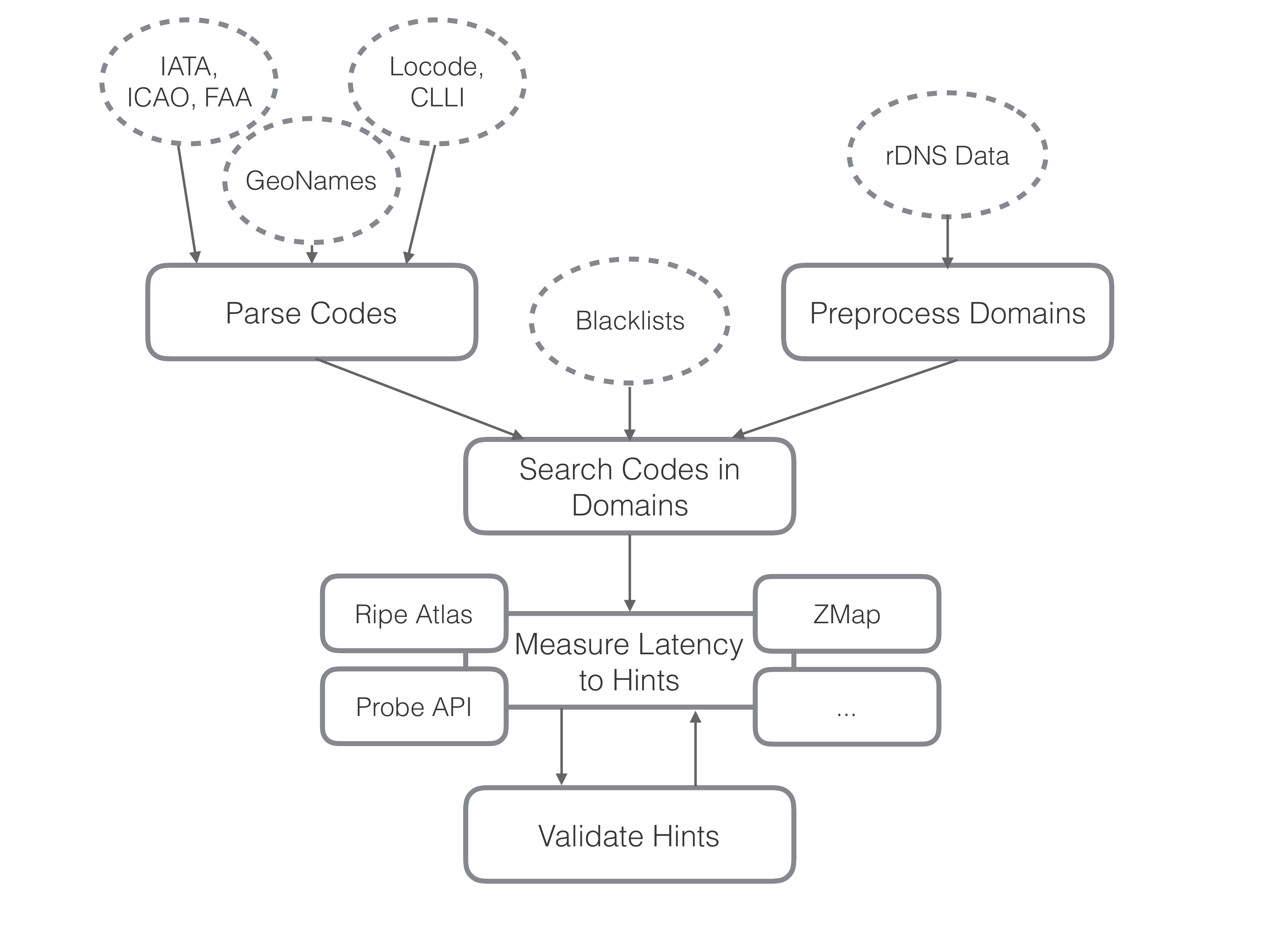}
	\caption{The building blocks of \hloc.}
	\label{fig:schema}
\end{figure}

HLOC comprises of five building blocks: (i) parse codes, (ii) preprocess domains (iii) search codes in domains (iv) measure latency to hints, and (v) validate hints. 
Figure \ref{fig:schema} shows these building blocks and their interfaces, which we describe in the following paragraphs.

\textbf{Parse Codes: }%
We gather several types of location codes, and map all possible \textit{codes} of one \textit{location} to a single location entry, amended by the location's coordinates and population data.
We use four kinds of location codes: Airport codes\cite{airportcodes} (example Houston, TX: IATA \texttt{HOU}, ICAO \texttt{KHOU}, FAA n/a), UN/Locode codes\cite{unece} (Houston: \texttt{US HOU}, transformed to \texttt{USHOU} for search), CLLI codes \cite{clli} (Houston: \texttt{HSTNTXMOCG0}, we use the first six characters \texttt{HSTNTX} as in \cite{huffaker2014drop}), and GeoNames city names (Houston: \textit{Hiustonas})\cite{geonames}.
The combination of those four kinds of locations codes results in 448k raw location codes.
These codes do not necessarily provide the same GPS coordinates for a location, especially for cities with rather distant airports.
We hence merge location codes in a radius of 100km around city centers to form one \textit{location}.
We use the number of inhabitants of a location to define a deterministic merge order, \ie smaller locations will be merged to bigger locations.
This results in 5474 locations.

\textbf{Preprocess Domains: }\label{sec:domainpre}%
\hloc's input is a list of target IP addresses with corresponding DNS names.
We use a reverse DNS file from Project Sonar \cite{sonarrdns} and filter it for routers using CAIDA's IPv4 ITDK \cite{caidaitdk} traces and IPv6 router names\cite{caidarouter} datasets.
We filter out $\approx$$1$\% of invalid domains, for example containing characters not allowed in domains according to RFC\,952 or top-level domains not existent according to IANA \cite{ianatlds}, for example, \textit{.local}.
For the remaining domains, we split DNS names into individual labels along ``.'' boundaries and remove first and second level domain labels (cf. Section \ref{sec:arch:sub:efficiency}).

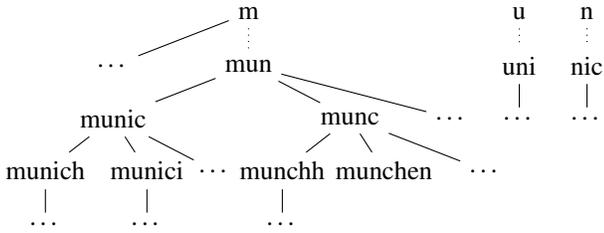
\begin{figure}
\begin{tikzpicture}[xscale=0.45,yscale=0.7]
    \path  (6,5) node(m) {m} 
		  (14,5) node(u) {u} 
		  (16,5) node(n) {n}
		  (6,4) node(mun) {mun}
		  (2,4) node(mun2) {\ldots}		  
          (12,3) node(munother) {\ldots}
          (2,3) node(munic) {munic}
          (5,2) node(municother) {\ldots}
          (0,2) node(munich) {munich}
          (0,1) node(munichother) {\ldots}
          (3,2) node(munici) {munici}
          (3,1) node(municiother) {\ldots}
          (9,3) node(munc) {munc}
          (13,2) node(muncother) {\ldots}
          (7,2) node(munchh) {munchh}
          (7,1) node(munchhother) {\ldots}
          (10,2) node(munchen) {munchen}
          (14,4) node(uni) {uni}
          (14,3) node(uniother) {\ldots}
          (16,4) node(nic) {nic}
          (16,3) node(nicother) {\ldots};
    \draw   (mun) -- (munic)
			(m) -- (mun2)
            (munic) -- (munich)
            (munic) -- (munici)
            (mun) -- (munc)
            (munc) -- (munchh)
            (munc) -- (munchen)
            (munic) -- (municother)
            (mun) -- (munother)
            (munich) -- (munichother)
            (munici) -- (municiother)
            (munc) -- (muncother)
            (munchh) -- (munchhother)
            (uni) -- (uniother)
            (nic) -- (nicother);
\draw[dotted]
   			(u) -- (uni)
   			(n) -- (nic)   			
   			(m) -- (mun);
\end{tikzpicture}
\caption{Prefix tree excerpt with relevant codes for search term \textit{munich}. Hierarchical layout for efficient search.}
\label{fig:trie}
\end{figure}

\textbf{Search Codes in Domains: }%
With 448k location codes and millions of domain names in preliminary runs, an efficient data structure and search algorithm is key to our research.
We find that organizing location codes in a prefix tree and then matching domains against this tree provides excellent performance, with about 10 minutes search time for our 1.6M domains (using 8 parallel processes).
Figure \ref{fig:trie} shows an excerpt of our location code prefix tree, with relevant codes for \textit{Munich}.
Matching a domain component \textit{munich} against our location code prefix tree would result in intermediate and leaf node matches for \textit{mun}, \textit{munic}, \textit{munich}.
As we also match the subcomponents \textit{unich}, \textit{nich}, \textit{ich}, the matches \textit{uni}, \textit{nic} and \textit{nich} are also generated.
Those matches of \textit{domain labels} against the prefix tree of \textit{location codes} are then named \textit{location hints}.

\textbf{Measure Latency to Hints: }%
\hloc is designed to flexibly use multiple measurement frameworks.
For our work, we implement interfaces to three servers running \zmap and to the \ripeatlas \cite{ripeatlas} measurement framework.
Extensions to use, e.g., PlanetLab\,\cite{planetlab} (currently missing IPv6 support) or ProbeAPI\,\cite{probeapi}, can  be flexibly added.
Queries and results are formed and processed in a unified format, which can easily be translated into framework-specific commands.
Per domain name, we start measurements against one \textit{location hint} and then call the \textit{validate hints} block.

\textbf{Validate Hints: }%
The \textit{validate hints} block is repeatedly called until all \textit{location hints} for a domain were \textit{falsified}, \ie  excluded based on speed-of-light constraints, or one \textit{location hint} was verified, \ie confirmed based on latency heuristics.

Specifically, \hloc's \textit{validate hints} block processes the set of responsive IP addresses by repeatedly (i) excluding locations by latency constraints and (ii) verifying remaining hints using \ripeatlas pin-point measurements. \\
Validation through pin-point measurements requires (a) finding a probe $p$ nearby  the \textit{location hint} $h$ (cf. Equation \ref{eq:dist})and (b) latency constraints (cf. Equation \ref{eq:lat}) holding true.
\begin{equation}\label{eq:dist}
d(p,h) < x
\end{equation}
\begin{equation}\label{eq:lat}
RTT(p,h) < a + \frac{2\cdot d(p,h)}{c \cdot c_0}%
\end{equation}
Equation\,\ref{eq:dist} limits the maximum distance of a probe $p$ to a location hint $h$.
For our study, we set $x$ to 1,000km, and highlight that \hloc will still choose the closest probe possible, but remote locations with sparse coverage benefit from a larger $x$. We discuss the impact of $x$ in Section \ref{sec:margins}.

In Equation \ref{eq:lat}, $RTT(p,h)$ is the measured round trip time between probe $p$ and hint location $h$, $a$ is a latency buffer, $d(p,h)$ is the geographical distance between probe and hint, $c_0$ the speed of light in vacuum, and $c$ the inverse refractive index for fiber. We use $c$$=$\nicefrac{2}{3}\,\cite{snyder2012optical}.

We argue that the latency buffer $a$ is necessary to accommodate for time a packet can spend in buffers or scheduling on shared measurement platforms.
Based on initial measurement results and our measurement targets, we consider $a$$=$9ms as a good starting value.
Studies requiring higher precision can adjust this parameter. We conduct a sensitivity analysis and discussion of this parameter choice in Section \ref{sec:margins}.

These settings result in a maximum possible error in our studies of $2x+a_{km} = \sim$2,900km, with $a_{km} = a{\cdot}c{\cdot}c_0\cdot{\nicefrac{1}{2}}$.

\subsection{Balancing Efficiency and Accuracy}
\label{sec:arch:sub:efficiency}
We resolve various challenges through iterative improvements, usually trading magnitudes of efficiency gain for minor potential losses in accuracy:

\textbf{Challenge 1 - Abundance of Matches: }%
Initial runs returned on average $>$$20$  mat\-ches per DNS name, resulting in an infeasible amount of matches. 
We contribute several approaches to reduce matches while not significantly reducing accuracy.

First, we focus on larger locations by removing locations below a certain inhabitant threshold. For this work, we choose a threshold of 100,000 inhabitants.
Again, other studies can easily chose a different threshold, and we conduct sensitivity analysis and discussion for this threshold in Section \ref{sec:sizethresh}.
This reduces the average number of matches per IP address from $>$$20$ to $\approx$$7$. It is possible to white-list important smaller locations (such as fiber landing points).

Second, we apply various blacklisting techniques to remove dominant false matches:
\begin{itemize}
\item 26 codes such as \textit{tel} (part of \textit{telecom}) or \textit{cpe} (\textit{customer premises equipment})
\item 425 words not to be searched for location hints, \eg \textit{Internet}, \textit{Linux} or \textit{static}
\item 35 code-location mappings considered wrong. For example, \textit{lin} is the IATA code for Milan, but also matches \textit{Illinois, Carolina} and \textit{Dublin}. In this example, the code-location blacklist will forbid \textit{lin} to match  \textit{Illinois, Carolina} and \textit{Dublin}.
\end{itemize}
We create these blacklists through iteratively assessing the top matches, thus focusing on high-impact false matches. 
This manual step may seem inefficient, but its one-off character and high-impact focus make it well feasible.
By filtering the 486 aforementioned values, we reduce the number of matches from $\approx$$7$ to $\approx$$4.4$. 
We thoroughly document our blacklist with examples to enable community review and contribution.

Third, we remove matches for top- and second-level domains.
These matches typically represent the headquarters' location (\eg \textit{.fr}) or the company name (\eg \textit{{vi\-ven\-di\-.fr}}), not specific router locations.
This brings the number of matches per IP address further down from $\approx$$4.4$ to $\approx$$3.9$.

Another option is to remove rDNS names that embed the IP address. These are typically  automatically generated without specific location hints. We support a variety of IPv4 and IPv6 embedding schemes. In this work, we separately investigate DNS names with and without embedded IP addresses. 

Through these measures we reduce the average number of matches per DNS name from $>$$20$ to $\approx$$3.9$.
Please note that this effect is amplified by fully removing IP addresses with no matches left, leading  to a similar reduction of total matches.

\textbf{Challenge 2 - Validation Runtime: }%
Measurement frameworks such as \ripeatlas allow a certain number of measurements per time interval, limiting the number of queries we can conduct.
Therefore, reducing the number of required measurements is our main lever to reduce validation runtime:
First, we precede the wide-spread measurements by conducting \zmap measurements from servers in Dallas, Frankfurt and Singapore, which allows speedy filtering of unresponsive IP addresses.
Also, the three geographically spread out servers can often locate the hemisphere of an IP address's location, excluding further matches.
This step brings the number of possible matches down from $\approx$$3.9$ to $\approx$$1.3$, and at the same time excludes unresponsive IP addresses.
%
\begin{table*}
	\caption{Statistics for Location Hints Matching.}
	\label{tab:routerstats}
	\centering
		{\begin{tabular}{lrrrrrrr}
				\toprule
				\# IP addresses & \multicolumn{3}{c}{IPv4} & \multicolumn{3}{c}{IPv6} \\
				\cmidrule(r){2-4}\cmidrule{5-7}
				Processing Step & Total & IP encoded & IP not encoded & Total & IP encoded & IP not encoded \\
				\midrule
				Routers & 2.5M (100\%) &  1.4M (100\%) & 1.0M (100\%) & 190k (100\%) & 146k (100\%) & 44k (100\%)\\
				Invalid Domains & -14k (0.6\%) &  \multicolumn{2}{c}{breakdown n/a}& -0.3k (0.1\%) &  \multicolumn{2}{c}{breakdown n/a}  \\
				No Match & -1.0M (\num{41.3}\%) & -821k (\num{58.2}\%) & -192k (\num{18.6}\%) & -7.2k (\num{3.8}\%) & -2.7k (1.8\%) & -4.5k (10.1\%) \\
				\midrule
				Remaining & 1.4M  (\num{58.2}\%) & 590k (\num{41.6}\%) & 836k (\num{81.6}\%) & 183k (\num{96.0}\% & 143k (98.2\%) & 39k (89.6\%) \\
				\bottomrule
			\end{tabular}}
\end{table*}

\begin{table*}
	\caption{Statistics for measurement algorithm: \zmap pre-scan effective for detection of unresponsive hosts, DNS names with no encoded IP addresses offer better results.}
	\label{tab:algostats}
	\centering
	{\begin{tabular}{lrrrrrrr}
				\toprule
				Processing Step & \multicolumn{3}{c}{IPv4} &~~& \multicolumn{3}{c}{IPv6} \\
				\cmidrule{2-4} \cmidrule{6-8}
				& Total & IP encoded & IP not encoded && Total & IP encoded & IP not encoded \\
				\midrule
				Input IP addresses & 1,426k ~(\num{100}\%) &~~590k~ (\num{100}\%) & 836k~ (\num{100}\%) && 183k~ (\num{100}\%) & \textbf{143k}~ (\num{100}\%) & 39k~ (\num{100}\%) \\
				 - Filtered IP addresses\textsuperscript{1} & -34k~~ (\num{2.4}\%) & -7k~~ (\num{1.2}\%) &-27k~~  (\num{3.3}\%) && -2k~~ (\num{1.3}\%) & -.2k~~ (\num{0.2}\%) & -2k~~ (\num{5.2}\%) \\
				- \zmap/censys timeouts & -391k (\num{27.5}\%) & -217k (\num{36.8}\%) & -174k (\num{20.8}\%) && -143k (\num{78.4}\%) & -139k (\num{97.1}\%) & -4k (\num{10.1}\%) \\
				- RIPE timeout & -40k~~ (\num{2.8}\%) & -22k~~ (\num{3.7}\%) & -18k~~ (\num{2.1}\%) && -8k~~ (\num{4.3}\%) &  -3k~~ (\num{2.3}\%) & -5k (\num{11.8}\%) \\
				Responsive & 961k~(\num{67.4}\%) &344k (\num{58.3}\%) & 617k (\num{73.8}\%) && 29k (\num{16.0}\%) & 1k~~ (\num{0.4}\%) & 29k (\num{73.0}\%) \\
				\midrule
				Responsive & 961k ~(\num{100}\%) & 344k~ (\num{100}\%) & 617k~ (\num{100}\%)&& 29k~ (\num{100}\%) & 1k~ (\num{100}\%) & 29k~ (\num{100}\%) \\
				~All hints falsified\textsuperscript{2} & 417k~ (\textbf{43.4\%}) & 132k (\num{38.3}\%) & 285k (\num{46.2}\%) && 7k (\num{22.9}\%) & .3k (\num{47.3}\%) & 6k (\num{22.4}\%) \\
				~\textbf{Hint verified} & \textbf{45k}~~~ (\num{4.7}\%) & \textbf{10k}~~ (\num{2.8}\%) & \textbf{35k}~~ (\num{5.7}\%) && \textbf{5k} (\num{17.6}\%) & \textbf{8}~~ (\num{1.3}\%) & \textbf{5k} (\num{18.0}\%) \\
				~No hint verified &500k~ (\num{52.0}\%) & 203k (\num{59.0}\%) & 297k (\num{48.1}\%) && 17k (\num{59.5}\%) & .3k (\num{51.4}\%) & 17k (\num{59.7}\%) \\
				\midrule
				~~Without probe & 17k~~~ (\num{1.8}\%) & 5k~~ (\num{1.5}\%) & 12k~~ (\num{2.0}\%) && 1k~~ (\num{4.2}\%) & .1k (\num{12.2}\%) & 1k~~ (\num{4.4}\%) \\
				~~Latency too high & 482k~ (\num{50.1}\%) & 197k (\num{57.4}\%) &284k (\num{46.1}\%) && 16k (\num{54.9}\%) & .2k (\num{39.3}\%) & 16k (\num{55.2}\%) \\
				\bottomrule
				\multicolumn{8}{l}{\footnotesize{1: Blacklisted or not announced IP addresses~~2: About 1\% falsified by \zmap/censys.}} \\
			\end{tabular}
		}
\end{table*}
\section{Measurement Results}\label{sec:measres}
In this section we present the results of our measurements in terms of verified and falsified \gl hints, differences between IP-encoding and not IP-encoding domain names, and contribution of each of our four location code types. %
\subsection{Measurement Statistics}
For our large-scale evaluation, we source IPv4 and IPv6 router IP addresses from CAIDA's May 2016 ITDK/DNS names datasets\,\cite{caidaitdk,caidarouter}.
From these, we filter out $\approx$$1$\% of invalid domains (see Section~\ref{sec:domainpre}).
The remaining domain names are then matched against the location hints in our prefix tree.
41\% (IPv4) and 4\% (IPv6) of IP addresses do not produce any location matches and are hence not further processed.
Some examples for such DNS names without matches are:
\textit{rf-rtr01.ew.net.nz},
\textit{host.3.static.cardbankph.com},
\textit{rt230bb131-145-61.routit.net} and
\textit{internet-gw.customer.alter.net}.
Table \ref{tab:routerstats} summarizes statistics for the aforementioned steps.
The set of IP addresses with matches (58\%/96\%) is then processed in the next steps of our algorithm.\\
{\noindent}We now describe the location measurement phase, with its statistics displayed in Table \ref{tab:algostats}.
We first filter 2\% of IP addresses from either our blacklist (curated from previous studies) or not announced to our BGP border router. 

Next, we conduct \zmap \textit{ICMP echo request} latency measurements and filter out unresponsive IP addresses.
These high-volume measurements eliminate 28\% of IPv4 addresses and 78\% of IPv6 addresses.
The value for IPv4 is in the expected range, given that our dataset also includes access and home routers.
The $78$$\%$ of non-responsive IPv6 addresses are surprising, however we find the total number of 29k responsive IPv6 routers roughly in line with previous studies (35k in \cite{Beverly2015ipv6}, 43k in \cite{Gasser2016ipv6}). %
Closer analysis reveals that a very large part~($75\%$) of our IPv6 dataset consists of dynamically generated IPv6 DNS names of the form \textit{{node-1w7jr9y4otrxqsxabcdek4c1.ipv6.telus.net}}.
This subset offers a response rate of only 0.4\%, which is not surprising as the addresses typically contained the SLAAC-characteristic \texttt{ff:fe} middle bytes.
These are therefore most likely home routers which could block ICMP requests.
Please note that the non-IP-encoded IPv6 subset offers a significantly higher ICMP response rate of 73\%.

Before discussing the breakdown of measured and responsive IP addresses into the three categories of \textit{verified hint}, \textit{no verified hint}, and \textit{all hints falsified}, we conduct a sensitivity analysis parameters of our algorithm.

\subsection{Sensitivity Analysis and Error Margins}\label{sec:margins}

We consider a \textit{location hint} as \textit{verified} if it satisfies Equations \ref{eq:dist} and \ref{eq:lat}.
These equations contain margins for probe distance and latency.
These margins are required, as finding a probe with zero distance to a location hint is as unlikely as a latency measurement equal to the direct distance.
Margin calibration strikes a balance between false positive and false negative rates.
Before discussing the results of our algorithm, we investigate the sensitivity of our results to these margins.

\begin{figure}
	\centering
	\includegraphics[width=\columnwidth]{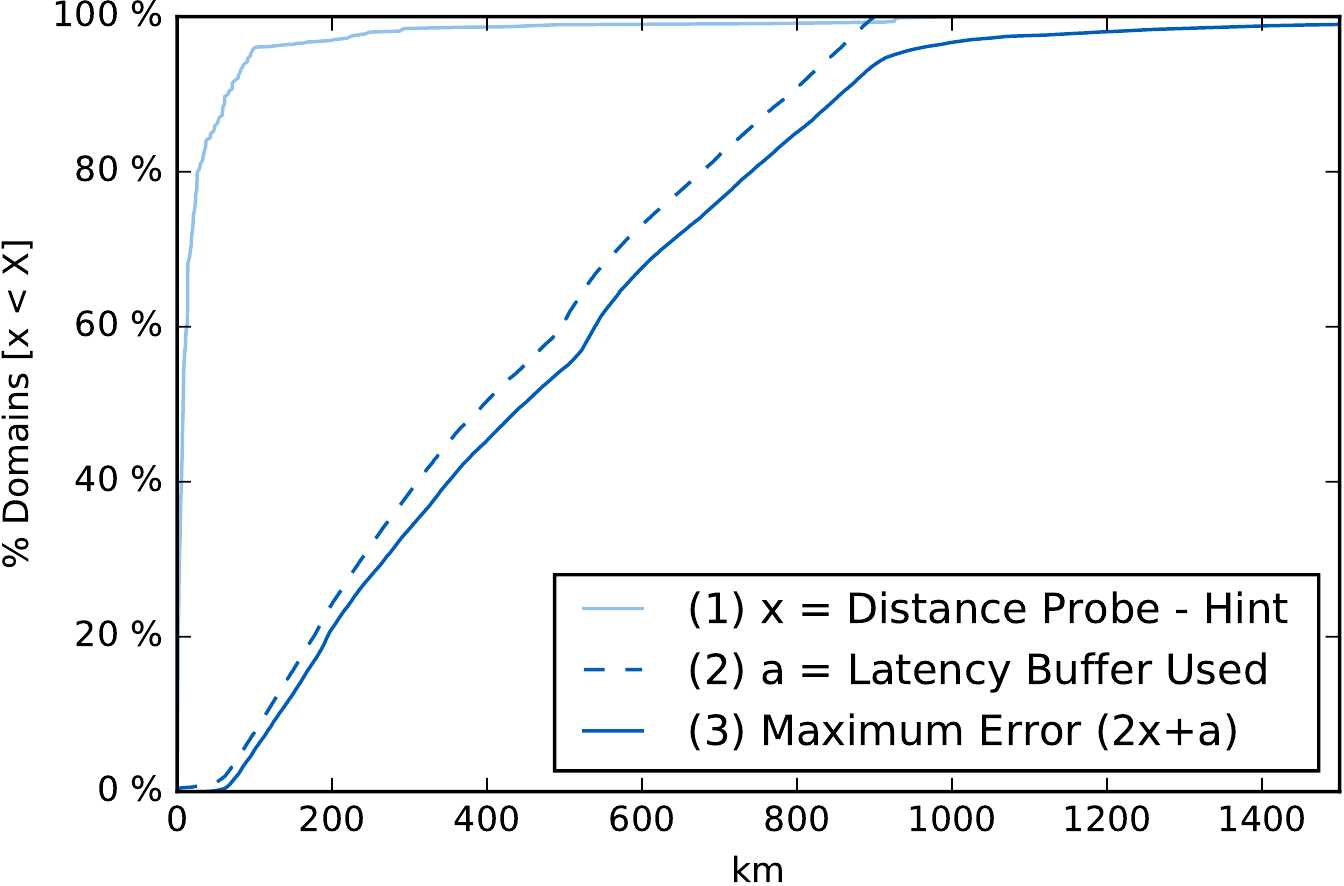}
	\caption{Verified location hints: CDF of (1) distance between probe and location hint -- 80\% under 25km, (2) Amount of latency buffer used - linear increase towards 9ms threshold, and (3) linear increase in maximum possible error, 93\% below 900km. IPv4 and IPv6 almost identical, hence mean shown.}
	\label{fig:v_rtts_cdf}
\end{figure}

Figure \ref{fig:v_rtts_cdf} investigates probe distance and maximum error for verified location hints. We see that probe distance is typically very small, with 80\% below 25km.
We also see that the latency buffer used, which we set to a maximum of 9ms, is rising linearly (\ie the underlying distribution is uniform in that interval).
The maximum error, as a function of both, also raises linearly to a point of 93\% of IP addresses at 900km.
We find the 7\% of high-error outliers to be typically caused by high probe distances. One such example is Summer Beaver, a remote airport in Canada, with the next \ripeatlas probe being 927km and a minimum of 16ms away.

\begin{figure}
	\includegraphics[width=\columnwidth]{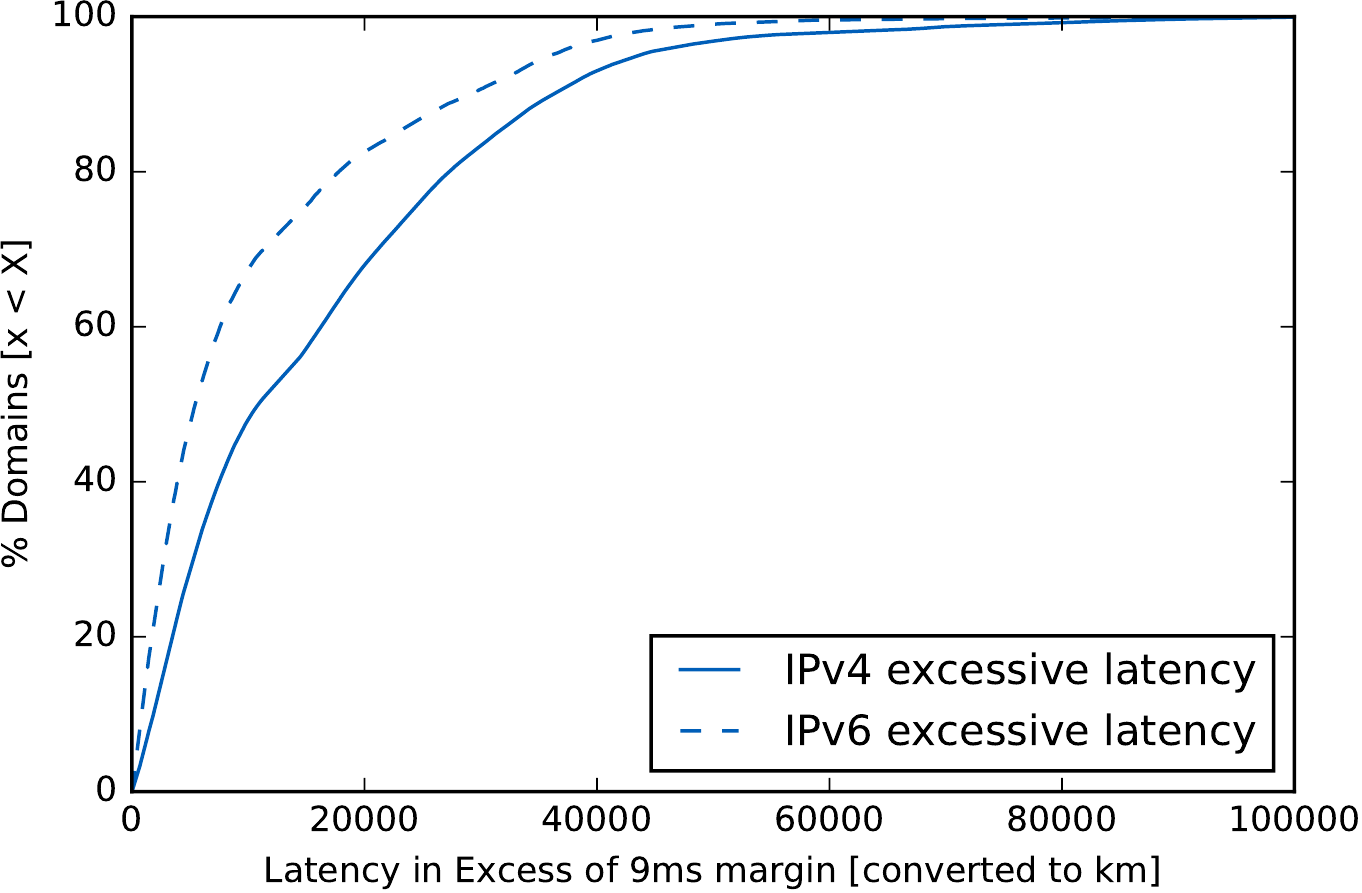}
	\caption{Not verified location hints: Excessive Latency (\rtt minus the verification threshold) raises linearly before dropping off into a long tail. IPv6 slightly better.}
	\label{fig:no_v_rtts_cdf}
\end{figure}

We next analyze the influence that our parameter choice has on not verified hints: Figure \ref{fig:no_v_rtts_cdf} analyzes the excessive latency of not verified hints, \ie how much a certain latency did exceed our thresholds for confirmation of a hint.
We see the excessive latency to continue the linear rise see in Figure \ref{fig:v_rtts_cdf} up to a point of $\sim$50ms, with slightly better values for IPv6. 

We conclude that, as the latency buffer used rises linearly until and beyond (cf. Figure \ref{fig:no_v_rtts_cdf}) 9ms, our selected margin of 9ms offers linear sensitivity to the number of verified hints and may easily be adapted by other studies to smaller or larger values.
For probe distance, permitting a maximum distance of 100km will permit to obtain 93\% of results without error margins rising beyond linear.
However, this would come at the cost of not being able to verify remote locations at all.

We next analyze the numbers of verified and unverified hints that result from applying our algorithm.

\subsection{Verified Hints}

We can verify a location hint for 4.7\% of responsive IPv4 (corresponds to 45k addresses) and 17.6\% of responsive IPv6 addresses (5k addresses).
While 4.7\% and 17.6\% are rather small portions of all initial IP addresses, we argue that 45k (IPv4) and 5k (IPv6) of successfully located router IP addresses prove that the concept of extracting DNS hints and validating them through \ripeatlas works for a significant number of routers.
We also find that the routers with a validated location play a central role in the Internet's topology:
Validated IP addresses were on average present in 12,585 of CAIDA's May 2016 path measurements compared to 3,567 traces for not validated IP addresses.
We argue that this more central role likely comes with higher uptime and response rates and lower latency, hence allowing for higher validation rates.
We discuss avenues to further raise the number of validated IP addresses in Section \ref{sec:disc}. %
\subsection{No Verified Hint}
We consider a domain as \textit{no verified hint} if we can neither verify any hint nor falsify all hints for a domain.
At about 55\%, the majority of responsive IP addresses falls into this category of \textit{no verified hint}.
One small root cause for this is the lack of a nearby probe, which amounts to 2\% (IPv4) and 4\% (IPv6).
This number of location hints without a nearby probe might be reduced through the adaption of further measurement frameworks and increased geographic coverage of \ripeatlas.
The majority of unverified IP addresses is due to high latency. We discuss mitigation strategies for this in Section \ref{sec:discussion}.
\subsection{All Hints Falsified}%
If all \textit{location hints} for a DNS name were falsified based on speed-of-light constraints, we label that DNS name with the results \textit{all hints falsified}.
With 43\% (IPv4) and 23\% (IPv6) of responsive IP addresses, this category has a surprisingly high share of all responsive IP addresses.
We manually investigate a sample of this group and find it to usually be caused by matches created on strings such as company names, which do not necessarily express an individual IP addresses location.
\subsection{IP-Encoding Domain Names}
This section compares the group of IP-encoding DNS names with that of not IP-encoding DNS names.
A common hypothesis is that IP-encoding DNS names are automatically generated, implying fewer or less accurate geographic hints. %

Tables \ref{tab:routerstats} and \ref{tab:algostats} split out this subgroup for comparison.
We find DNS names with encoded IP addresses to represent a significant share of IPv4 (41\%)  and IPv6 (78\%) router IP addresses.
Those DNS names that encode IP addresses generate fewer location matches, respond less likely to \textit{ICMP echo requests} and less likely have their location hints validated.

However, 10k IPv4 addresses from the IP-encoding provide verified location hints.
We further analyze this verified subgroup IP-encoding DNS names, and commonly find city or regional names in labels above the IP-encoded label.
Table \ref{tab:ipenc-but-match} displays examples of this behavior. %
\begin{table}[h]
	\caption{Examples of verified IP-encoding DNS names.}%
	\label{tab:ipenc-but-match}
	\resizebox{\columnwidth}{!}{
		\begin{tabular}{llcr}
			\toprule
			DNS Name & Location & Code & RTT \\
			\midrule
			\textit{\footnotesize{1-2-3-4.lightspeed.\textbf{hstntx}.sbcglobal.net}} & Houston, TX & CLLI & 3.5ms \\
			\textit{\footnotesize{1-2-3-4.lightspeed.\textbf{miamfl}.sbcglobal.net}}. & Miami, FL & CLLI & 4.2ms \\
			\textit{\footnotesize{ip-1-2-3-4.\textbf{mel}.xi.com.au}}. & Melbourne (AU)& IATA & 6.0ms \\
			\textit{\footnotesize{rrcs-1-2-3-4.\textbf{nyc}.biz.rr.co}}. & NYC, NY & IATA & 3.7ms \\
			\bottomrule
		\end{tabular}%
	}%
\end{table}%
\subsection{Contributions per Location Hint Source}\label{sec:contribsource}
We next evaluate the contribution of various location sources to the number of generated and verified location hints.
Table \ref{tab:sourcestats} shows location hint sources along their numbers of codes, generated hints, and verified hints.
Over 99\% of verified matches stem from IATA, Geonames and CLLI codes, while ICAO, FAA and UN/LOCODE codes provide very few verified hints.
GeoNames and CLLI matches are very efficient when comparing number of hints to number of verified hints.
As CLLI and GeoNames codes are longer than, \eg IATA's 3-letter codes, they are less prone to matching strings by chance.
We discuss in Section \ref{sec:disc} how these insights can potentially improve \hloc by weighting location hint sources.
\begin{table}[h]
	\caption{IATA, GeoNames and CLLI codes provide 99\% of verified hints.}
	\label{tab:sourcestats}
	\resizebox{\columnwidth}{!}{
		\begin{tabular}{lrrrrrr}
				\toprule
				Category & IATA & ICAO & FAA & UN/LO & GeoNames & CLLI\\
				\midrule
				\# Codes & 8k	& 13k& 20k & 77k & 32k & 31k \\
				\midrule
				Hints (100\%)	 & 4.5M & 209k & 472k & 59k & 215k & 167k \\
				Verified & \textbf{32k} & 122 & 413 & 120 & \textbf{13k} & \textbf{5k} \\
				Verified (\%) & .7\% & $<.0$\% & .1\% & $<.0$\% & \textbf{5.9\%} & \textbf{2.8\%}\\
				\bottomrule
			\end{tabular}
		}
	\end{table}

%
\section{Evaluation}\label{sec:eval}

In this section we compare the locations obtained from our \hloc \gl framework to (i) commercial \gl databases and (ii) to Huffaker et al.'s results obtained by using their \drop system \cite{huffaker2014drop}. %
Table \ref{tab:evalstats} details this comparison.
\begin{table*}[tb]
	\caption{Evaluation of location decisions by databases and \drop against \hloc measurements: ip2location more accurate than GeoLite, \drop frequently with ``no data''. All information-based approaches with a significant number of wrong decisions.}
	\label{tab:evalstats}
		\centering
		{\begin{tabular}{clrrrrrrrrrrrrrrrr}
				\toprule
                & \multicolumn{2}{c}{\hloc} && \multicolumn{3}{c}{GeoLite} && \multicolumn{3}{c}{ip2location} && \multicolumn{4}{c}{\drop} \\
				\cmidrule{1-3} \cmidrule{5-7} \cmidrule{9-11} \cmidrule{13-16}
                & Location Dec. & $n$ && Same & Possible & Wrong && Same & Poss. & Wrong && Same & Poss. & Wrong & No data \\
				\midrule
                \multirow{4}{*}{\rotatebox[origin=c]{90}{IPv4}}
                & Verified & 45k && \num{40.4}\% & \num{15.6}\% & \textbf{44.0\%} && \textbf{76.6\%} & \num{11.3}\% & \textbf{12.1\%} && \num{7.8}\% & \num{.1}\% & \num{8.4}\% & \textbf{83.7\%} \\
                & All falsified & 417k && n/a\textsuperscript{1} & 100\% & 0\% & & n/a & 100\% & 0\% && n/a & n/a & 2.2\% & 97.8\% \\
                & No verified~~ & 499k && n/a & 96.1\% & 3.9\% & & n/a & 98.8\% & 1.2\% && n/a & 10.5\% & 4.1\% & 85.4\% \\
                & Timeout & 465k && n/a & 100\% & n/a\textsuperscript{2} & & n/a & 100\% & n/a && n/a & 26.4\% & n/a & 73.6\% \\
				\midrule
                \multirow{2}{*}{\rotatebox[origin=c]{90}{IPv6}}
                & Verified & 5k && \multicolumn{3}{c}{---} && 25.7\% & 10.6\%  & \textbf{63.6\%} && 33.7\% & 1.0\% & 1.8\% & 63.5\%\\
                & No verified & 17k &&  \multicolumn{3}{c}{---} && n/a & 74.2\% & \textbf{23.9\%} && n/a & 25.5\% &  3.3\% & 71.2\% \\
				\bottomrule
				\multicolumn{16}{l}{\scriptsize{1: With no verified HLOC match, other approaches can not have the same match.~~2: With HLOC timeout, it is not possible to evaluate other approaches.}}%
			\end{tabular}
		}
\end{table*}

\subsection{Comparison to commercial databases}
We compare our results against MaxMind's\,\cite{geolite2} and ip2location's city-level \gl databases\,\cite{ip2location}. 
We report findings in three possible categories: Based on \hloc's hints and measurements, a database entry may either be the \textit{same}, \ie \hloc could verify a location hint and the database reports the same location, \textit{possible}, \ie the  database location is different from \hloc, but possible based on \hloc's latency measurements, or \textit{wrong}, \ie the location reported by a database has been falsified based on \hloc's latency measurements and speed-of-light constraints. 
\\
\textbf{\hloc Verified IPv4: }%
We first look at IPv4 locations that have been verified by \hloc.
Looking at the \textit{verified (IPv4)} row in Table \ref{tab:evalstats}, we see that, based on our measurements, 44\% of locations reported by GeoLite are wrong, \ie falsified through measurement, compared to about 12\% from ip2location.
This confirms that commercial databases, when applied to router datasets, contain a non-negligible number of incorrect entries.
In addition, we can compare the percentage of cases where a verified \hloc location is the same as the city reported by a database. 
We highlight that as neither the databases nor \hloc form a definitive ground truth, the \textit{same} category is rather indicative. However, the difference of 77\% agreement for ip2location and 40\% for GeoLite is noteworthy.
\textbf{\hloc Unverified IPv4:}
For location hints that \hloc can not verify, \hloc's latency measurements still disprove a small number of locations returned by commercial databases. 

\textbf{IPv6: }%
Table \ref{tab:evalstats} displays IPv6 results where they significantly differ from IPv4 results. 
GeoLite does not support IPv6, and ip2location IPv6 matches are significantly more often wrong than IPv4 matches (64\% IPv6; 12\% IPv4). 
This is likely linked to the large IPv6 address space, pressuring database providers to aggressively group IP addresses into the blocks mentioned in Section \ref{sec:relwork}. 
This highlights the need of contrasting database information with latency-based measurements  for IPv6.

\subsection{Comparison to \drop's ruleset}

\drop does not provide a \gl for 84\% of IPv4 router addresses. 
This is caused by \drop's operating mechanism based on specific regular expression rules for a set of specific domains.
However, for those IP addresses where \drop does deliver an answer, the answer interestingly is typically verified or falsified, and only very rarely in the ``possible'' range. 
We highlight that although the \drop rules are very narrow, specifying a specific kind of code (for example, IATA) at a specific location in a specific domain, its answers are still frequently falsified through latency measurements. 
This confirms the work of Zhang et al.\,\cite{zhang2006misnaming} stating that DNS names are often outdated or wrong. 
It also confirms that DNS-based location approaches should be combined with latency measurements to minimize the influence of wrongly named IP addresses.
Interestingly, the \drop results are much better for IPv6 - this might be due to less pressure to re-use and re-locate addresses in the vast IPv6 address space.

\subsection{Comparison to \drop ground truth domains}
\begin{table*}[t]
	\caption{For \drop's ground truth domains, we show performance for (a) \drop's reported performance, (b) our reproduction of \drop and its validation against latency measurements, and (c) \hloc-generated hints and their latency validation.}
	\label{tab:dropgt}
	\begin{minipage}{\textwidth}
		\resizebox{\textwidth}{!}
		{\begin{tabular}{lrrrrrrrrrrrrrrrrrr}
				\toprule
				Domain &  \multicolumn{4}{c}{\drop 2014 \cite{huffaker2014drop}} && \multicolumn{5}{c}{\drop 2016 Reproduction} && \multicolumn{4}{c}{\hloc}  \\
				\cmidrule{2-5}\cmidrule{7-11}\cmidrule{13-16}
							&	$n$ & Type & Match & TP\textsuperscript{1} && $n$ & Match & TP\textsuperscript{1} & Ver.\textsuperscript{2}&  Fals.\textsuperscript{2} && Match& TP\textsuperscript{1} &Ver.\textsuperscript{2} & Fals.\textsuperscript{2} \\
				\midrule
				belwue.de & \num{161} & City  & 52\% & 86\% && \num{53} & 64\% & 65\%& 22 &   1 & & 94\% & 64\% & 32 &  5  \\ 
				cogentco.com & \num{13129} & IATA & 90\% & 99\% && \num{9475} & 95\% & 26\% & \num{2381} & \textbf{628} & & 99\% & 23\% & \num{2144} & 295 \\ 
				digitalwest.net & \num{111} & IATA & 49\% & 100\% && \num{47} & 49\% & 26\% & 6 &   0 & & 100\% & 15\%&  7 & 2 \\
				ntt.net & \num{2584}  & CLLI & 96\% & 100\% && \num{3125} & 54\% & 37\% & \textbf{622} &   5  && 99\% & 30\% & \textbf{937} & 148\\ 
				peak10.net & \num{115} & IATA & 100\% & 100\% && \num{199} & 99\% & 9\% & 18 &  0  && 100\% & 9\% & 18 & 0\\ 
				\bottomrule
				\multicolumn{16}{l}{\scriptsize{1: \% of matches that are true positives~~2: Total count of verified or falsified matches. ``possible'' and ``time out'' results not displayed.}} \\
			\end{tabular}}
		\end{minipage}
\vspace{-5mm}
\end{table*}
In this section, we compare the accuracy of \hloc and \drop against the set of 5 ground truth domains from \drop. 
For those domains, the authors of \drop had confirmed the accuracy of their rule set. \\
Table \ref{tab:dropgt} first shows the original results of Huffaker et al. \cite{huffaker2014drop}.
As they do not provide their dataset, we reproduce their results for a host-by-host comparison, which we show next. 
Please note the different number of hosts $n$, as we filter for all routers in a domain (according to CAIDA traces), while \drop authors likely picked the subset of hosts that they obtained a ground truth for.
Also, we limit this comparison to IPv4 as original \drop numbers do not include IPv6. 
We reproduce \drop results by applying their published matching rules and run \hloc's measurement algorithm to verify or falsify their results. 
We also give the results for matches generated and measured by \hloc.
Our reproduction, though working on a different numeric base, typically achieves similar match rates. 
The difference between the original \drop result and our reproduction of \textit{ntt.net} is likely caused by NTT's creation of international CLLI codes, which are not part of the standard and are hence missing in our dataset. 
The \drop team had likely created these easily decipherable codes such as \textit{londen}, \textit{taiptw}, \textit{newthk} and \textit{amstnl} by hand. 
Using \hloc, we can typically verify a large part of the \drop reproduction matches, but for \textit{{cogentco.com}} can also falsify over 600 matches. 
The last aggregate column are the results of \hloc. 

The application of \hloc's full matching tree generates more matches than the \drop rules, with the count of verified matches frequently equaling or surpassing the number of verified matches from \drop's rules.
We investigate those cases where \hloc can verify matches that the \drop rules can not find. 
We find no systematic pattern, but incidences of (i) manually named DNS records not in accordance with \drop rules and (ii) finding (and verifying) matches in DNS names that seem not to intend to give a location hint. 
Based on our reproduction of \drop, we argue that \hloc's generic matching equals or surpasses \drop's specific domain patterns just two years after their creation.
We also highlight that even specific \drop rules can be falsified for a significant number of hosts. 
This confirms the unavoidable existence of outdated and misnamed DNS names, requiring latency based verification/falsification as done by \hloc.
\section{Discussion and Limitations}\label{sec:disc}\label{sec:discussion}
We discuss our work, its current limitations, and ways to tackle these in this section. 

Throughout this work, we have confirmed that for router datasets, \gl database results have a significant number of wrong entries and should be contrasted by latency measurements, an issue even worse with IPv6. 
We show that leveraging ready-to-use public measurement frameworks in \hloc works well, and that a significant number of \textit{location hints} can be confirmed with singular latency measurements.
Using few and simple measurements is an important scale factor when leveraging public frameworks that often charge by amount and complexity of measurements. 
We find the error margin to increase linearly for a very large fraction of measurements, easily enabling users of \hloc to set an individually acceptable error margin.

We now discuss \hloc's current design choices and plans for improved future versions.

\subsection{Measurement Scope and Nature}

Since Internet-wide geolocation measurements from hundreds of distributed probes are infeasible, we limit \hloc's scope to geolocating routers.
This also ensures that we adhere to \ripeatlas API thresholds, since \hloc is measurement-bound and not computationally-bound.
In the future we want to perform multiple measurements for one router over time to avoid temporary congestion and queuing situations.

\subsection{Probe Selection}\label{sec:probeselection}
The current version of \hloc randomly selects a \ripeatlas probe close to a location hint. 
This offers various ways for improvement: 
First, \ripeatlas probes with high first- or second-hop latencies could be removed.
As each \ripeatlas probe continuously measures first- and second-hop latency, probes behind high-latency links can be excluded from our measurements. 
Second, selection of probes within the same Autonomous System as the target IP address can avoid high latencies due to geographically sparse inter-AS connections.
Third, we could select several suitable probes, aiming to increase the probability of hitting a low-latency measurement caused by an empty link or a low probe workload. 
Holterbach et al.~\cite{holterbach2015quantifying} show that measurement interference can significantly bias \ripeatlas latency measurements.
This step increases the measurement cost and might not be feasible for some studies.
Fourth, consensus-based blacklisting can remove probes with wrong locations.
Fifth, the integration of further frameworks can aid probe sparsity and probe diversity.

\subsection{Probe Distance, Latency Buffer, and Error Margin}\label{sec:latbuf}
As discussed in Section \ref{sec:margins}, our parameters for probe distance and latency buffer offer largely linear sensitivity and can easily be adjusted to suit different requirements. 
A smart way to combine both parameters will be for the user to set a maximum acceptable error margin, towards which \hloc will optimize by dispatching more measurements for extreme cases.
Future versions may also show possible location areas of an IP address, however these will be of complex shape.

\subsection{Location Codes Ordinance and Search}\label{sec:ord}
Currently, \hloc probes matches for a domain in the order in which they are found in our code prefix tree.
Future versions should leverage smart sorting to reduce the average number of measurements needed to come to a conclusion for a domain.
Supported by our results from Section \ref{sec:contribsource}, this could include sorting by match length (putting higher weight on longer matches), by confirmation count (\ie measuring frequently correct hints first), or by the location of a match within a label (e.g., prioritizing isolated matches over matches within a word). Reinforcement learning to determine typical code types and locations per domain \cite{huffaker2014drop} may also help.

\subsection{Location Size Threshold}\label{sec:sizethresh}
The current version of \hloc uses a location size threshold of 100k inhabitants: 
Only locations above this threshold will be able to generate matches.
This threshold is mainly applied to reduce the abundance of matches generated per domain. 
As discussed in Section \ref{sec:arch}, this threshold can both be changed for its absolute value as well as for individual exceptions.

We conduct a sensitivity analysis on this threshold value, evaluating the average number of matches per domain after exclusion of locations based on \zmap latency measurements. 

Using a threshold of 100k inhabitants yields $7.7$ average matches per domain (cf. Section \ref{sec:arch}). 
For a threshold of 50k, this rises to $11.34$, and for a threshold of 1k, this number rises to even $20.44$.
As our measurements already exceed many of \ripeatlas's limitations for a threshold of 100k, we conclude that 100k was the correct threshold to choose for our large-scale measurements. 
Typical path analysis use cases \cite{schmitt2016mvno, scheitle2016analyzing}, only locate thousands instead of millions of IP addresses and could easily use a lower threshold.
Also, the steps discussed in Section \ref{sec:ord} could likely reduce the number of required measurements for domains with many matches. 

\subsection{Measuring IP Anycast Addresses}

IP anycast routes packets for one IP address to one of several distinct nodes, usually in separate geographical locations, used for example by DNS root servers \cite{dnsroot}.
When performing latency measurements, \hloc selects a probe geographically close to each hint.
As \hloc, in its current version, stops after having validated a first hint, it would disprove other valid locations for multicast IP addresses. 
This has a small impact on our results, as (i) IP anycast addresses typically do not encode locations in their DNS name; and (ii)  we are not aware of IP anycast being used for Internet routers to a significant extent.

\subsection{Ethical Considerations}\label{sec:ethics}%
We follow an internal multi-party approval process, among others based on Partridge and Allman~\cite{partridge2016ethical}, before any measurement activities are carried out. 
We conclude that our \textit{icmp echo request} measurements consisting of a few packets per week per targeted router, and the resulting data, can not harm individuals, but may result in investigative effort for system administrators.
We aim to minimize this effort by deploying scanning best-practice efforts of (i) using dedicated scan machines with explanatory websites, (ii) maintaining a blacklist, (iii) replying to every abuse e-mail (none received in this experiment), and (iv) minimizing impact on \ripeatlas in coordination with the \ripeatlas team. 
\section{Conclusion}\label{sec:concl}
We present \hloc, a framework that derives geolocation hints from DNS names and uses publicly available measurement frameworks to validate or falsify theses hints through latency measurements. 
We evaluate its performance on IPv4 and IPv6 router datasets and verify 45k IPv4 addresses and 5k IPv6 addresses. 
We compare our results to those from \gl databases and \drop and find that we can frequently disprove those base on latency constraints.
We provide \hloc as ready-to-use tool for other research groups.

\noindent\textbf{Data and Code Release: }%
In~\cite{reproduc2017}, we outline our aim for repeatable, replicable and reproducible research as defined by ACM~\cite{acmrep}, and publish \hloc code and data on GitHub: 
\centerline{\texttt{\url{https://github.com/tumi8/hloc}}}
This repository will also be used for future development. 

\noindent\textbf{Acknowledgments: } We thank Johannes Naab for his input, the  Leibniz  Supercomputing  Centre  (LRZ)  of  the  Bavarian Academy of Sciences (BAdW) for cloud computing infrastructure, and the \ripeatlas team for their support. 
This work has been supported by the German Federal Ministry of Education and Research, project X-CHECK, grant 16KIS0530, and project AutoMon, grant 16KIS0411.
%
\apptocmd{\sloppy}{\hbadness 10000\relax}{}{}
\apptocmd{\thebibliography}{\raggedright}{}{}
\bibliography{geoloc}
\bibliographystyle{abbrv}
\end{document}